# • Prediction of Stable Cu-Li Binary Intermetallics From First-Principles Calculations: Stoichiometries, Crystal Structures, and Physical Properties

Jia Hui Yu[2], Dawei Zhou[1*], Xin Tang[3], Chun Ying Pu[1,*]

[1] *College of Physics and Electronic Engineering, Nanyang Normal University, Nanyang 473061, China*

[2] *School of Electronic and Electrical Engineering, Nanyang Institute of Technology, Nanyang 473004, China*

[3] *College of Material Science and Engineering, Guilin University of Technology, Guilin 541004, China*

## Abstract:

Towards a resolution of the longstanding controversy regarding the existence of Cu-Li intermetallic compounds, we extensively investigate the phase stability of Cu-Li intermetallics with various possible stoichiometries at zero temperature and pressure using a global structure searching method. It is found that Cu-Li intermetallics can exist stably at atmospheric pressure, and three stable intermetallics (*Fmmm* $Cu_1Li_2$, $Fd\bar{3}m$ $Cu_2Li_1$ and $P\bar{1}$ $Cu_7Li_1$,) are identified. Electronic structure analysis reveals that although the three stable phases are metallic, covalent Cu-Cu and ionic Cu-Li bonds are found in the three structures. Moreover, the 3d states of copper atoms are mostly responsible for bond formations in the Cu-Li intermetallics. For all the predicted Cu-Li intermetallics, the effect of Cu concentration on structure, mechanical and thermodynamic properties are calculated systematically. It is found that the copper atoms in Cu-Li intermetallics trend to form covalent bonds, so more covalent bonds are formed as Cu content increases, leading to the increases in the elastic moduli, Vicker hardness and Debye temperature with Cu content on the whole. The Poisson's ratios of Cu-Li intermetallics vary in the range of 0.25 and 0.35, and most of Cu-Li intermetallics exhibit a good ductile property. The elastic anisotropy calculations suggest that all the Cu-Li intermetallics show anisotropic elasticity more or less, and the percentage anisotropy in compressibility is smaller than that in shear for each of the predicted Cu-Li compounds.

* Corresponding author email: zhoudawei@nynu.edu.cn and puchunying@126.com

**Key words:** Phase diagram; Intermetallics; First principle; Thermodynamic properties

## 1. introduction

Copper is an abundant, fairly inexpensive transition metal. Copper and its alloys have been widely used in electric industry because of their excellent electrical conductivity. However, the impurities in Cu usually cause a decrease in electrical conductivity. It has been revealed that lithium can not only refine the impurities in copper, but also increases the tensile strength of copper without the electrical conductivity decreasing [1]. However, there remains a great deal of controversy regarding the existence of Cu-Li intermediate phases over past eighty years. Both Pastorello [2] in 1930 and Klemm *et al*. [3] in 1958 showed that there are no intermediate phases in the Cu-Li phase diagram through X-ray analysis. Pelton [4] also believed that there are no intermediate phases in binary Cu-Li alloys based on the thermodynamic calculations. Saunders [5] further supported the nonexistence of Cu-Li intermetallic phases in 1998. On the contrary, Old and Treven [6] proposed a cubic $Cu_4Li$ intermediate phase based on XRD investigations as early as 1981. Later, Penaloza et al. [7] and Borgstedt [8] accepted the existence of $Cu_4Li$. Van de Walle *et al*.[9] also suggested that $Cu_4Li$ phase could exist probably based on *ab-initio* calculations. Besides $Cu_4Li$, Gąsior *et al*. [10] in 2009 proposed an intermetallic phase $Cu_2Li_3$ by means of electromotive force measurements, but they didn't provide clear evidence. Okamoto [11] further accepted the existence of $Cu_2Li_3$ reported by Gąsior *et al*. [10].

Very recently, D. Li *et al*. [12] pointed out that the lattice parameters of the $Cu_4Li$ phase proposed by Old and Treven *et al* [6] fit well with the lattice parameters of fcc Cu, therefore, $Cu_4Li$ phase was considered to be part of fcc Cu, they also declared that they could not confirm the presence of $Cu_2Li_3$ using the powder XRD measurements. We noticed that the previous work [13-14] have showed that Au and Li can form stable intermetallic phases. It is all known that elements in the same group often have similar physical and chemical properties. Therefore, it seems possible for Cu and Li to form stable intermetallic phases in theory.

As we known, first-principle calculations are based on the laws of quantum mechanics, which have been successfully applied to the investigations on the physical properties of alloy, such as phase diagram, electron structure, phonon curves, mechanical and thermodynamic properties[15-18]. So in this paper, we systematically explored the ground –state phases of Cu–Li system using evolutionary structure prediction algorithm USPEX and first principle calculations. We found that the stable Cu-Li intermetallics can be formed under normal conditions, and three stable Cu–Li intermetallic compounds, *Fmmm* $Cu_1Li_2$, $Fd\bar{3}m$ $Cu_2Li_1$ and $P\bar{1}$ $Cu_7Li_1$, were found. We further investigated systematically the physical properties of all the predicted ordered Cu-Li alloys as a function of Cu concentration, including the structure stability, formation enthalpy, phonon spectrum, electronic, mechanical and thermodynamic properties.

## 2. **Theoretical method**

In this work, structure searches for the stable Cu-Li intermetallic phases have been performed across the entire concentration range using the evolutionary algorithm-based method Universal Structure Predictor: Evolutionary Xtalloraphy (USPEX) [19,20], which has been successfully applied to predict structures of many systems from elemental solids to ternary compounds [21-27]. The structure searches were carried out up to 20 atoms in the unit cell for all possible stoichiometries. In each search, the first generation of structures was produced randomly and subsequently optimized. Each generation contains 30 structures. For the next generation, 60% of the structures were generated from the lowest-enthalpy structures provided by the previous generation, while 40% would be generated randomly. We usually followed 20 generations to achieve convergence of the sampling of the low-energy minima in configurational space. All the structural relaxations were performed within the framework of density functional theory using the Vienna *ab initio* Simulation Package[28,29]with the projector augmented wave (PAW) method [30]. The first-principles pseudo-potential Perdew-Burke-Ernzerhof (PBE) approach was chosen as the exchange and correlation functional[31]. The PAW potentials treated $3d^{10}4s^1$ and $2s^1$ as the valence electrons for the Cu and Li, respectively. For structure prediction, a plane-wave kinetic energy cutoff of 600 eV and a k-point grid spacing of $2\pi\times0.06$ Å$^{-1}$ were used. When we calculated the enthalpy curves and band structures, a higher level of accuracy consisting of a basis set cutoff of 600 eV and a k-point grid spacing of $2\pi \times 0.03$ Å$^{-1}$ were used, which ensured that those calculations were well-converged. During the structural optimization, all atoms were fully relaxed until the total energy changes and force less than $1.0\times10^{-6}$ eV/atom, 0.002 eV/Å, respectively. The phonon dispersion calculations were performed by using a supercell approach as implemented in the PHONOPY code [32,33]. The phonon calculations for $Cu_1Li_2$, $Cu_2Li_1$and $Cu_7Li_1$ were performed by using $1\times1\times2$, $2\times2\times2$ and $2\times2\times2$ supercells, respectively.

## 3. Results and discussions

### 3.1. USPEX searching the stable structures

Our structure searches successfully reproduced the experimentally observed fcc Cu and bcc Li at zero pressure, and the calculated lattice parameters agree well with experimental results [34] as shown in **Table S1**. In order to describe the thermodynamical stability of Cu-Li alloys, we calculated their formation enthalpies. The enthalpy of formation ΔH can be calculated from the following relation:

$$\Delta H = [E(\mathrm{Cu}_m\,\mathrm{Li}_n) - mE(\mathrm{Solid\ Cu}) - nE(\mathrm{Solid\ Li})]/(m + n) \qquad (1)$$

where $\Delta H$ is defined as the relative formation enthalpy per atom of a compound of this stoichiometry, and E($Cu_mLi_n$) is the total energy/ formula unit of the compound, E(Solid Cu) and E(Solid Li) are the total energies of the pure elements in their most stable state, and *m* and *n* are the number of Cu and Li atoms for a system, respectively. In general, a negative formation enthalpy means that the crystal can exist stably. Moreover, the more negative formation enthalpy indicates, the higher stability of crystal. In Fig.1, the convex hull connects the phases with the lowest formation enthalpies among all compositions, and any phases lying exactly on the convex hull are deemed as energetically stable. The structures whose enthalpies remain above the convex hull would be thermodynamically metastable. From Fig.1, we can conclude that stable intermetallic phases of Cu-Li alloys can be formed at zero pressure due to a negative enthalpy of formation from Cu and Li. For the considered Cu-Li compounds, three stable structures are found, i.e., $Cu_1Li_2$, $Cu_2Li_1$ and $Cu_7Li_1$. The schematic illustrations of them are shown in Fig.2. The corresponding lattice parameters and atomic positions are listed in **Table 1**. $Cu_1Li_2$ crystalizes in an orthorhombic structure with space group *Fmmm*. From a structural point of view, *Fmmm* $Cu_1Li_2$ can be regarded as a layered structure. In this structure, a isosceles triangle unit consisted of three Cu atoms is found, and many such units form a coplanar Cu layer, while Li atoms form two puckered layers between the two Cu-layers. The Cu-Cu bonds distances in $Cu_1Li_2$ are 2.475 and 2.505 Å, respectively, which are shorter than those in fcc Cu (2.556 Å), so there may exist a stronger interrelation among the Cu atoms in $Cu_1Li_2$. The most stable stoichiometry is $Cu_2Li_1$, which has a cubic structure with space group $Fd\bar{3}m$. Interestingly, Li atoms simply form a diamond lattice in this structure, while Cu atoms form many tetrahedron units embedded in the lithium framework. All the lengths of Cu-Cu bonds in the tetrahedron unit are 2.454 Å, which are also shorter than those in fcc Cu. It's important to note that there are some Cu-Li intermetallics with high copper content such as $Cu_5Li_1$, $Cu_6Li_1$, $Cu_7Li_1$, $Cu_8Li_1$, and so on.. Among these high copper content compounds, $Cu_7Li_1$ is a thermodynamically stable phase, while the other phases are slightly above the convex hull and might exist as metastable phases. The stable $Cu_7Li_1$ has a triclinic structure with space group $P\bar{1}$, the Cu atoms in which also form tetrahedron units or planar triangle units, while lithium atoms are embedded in the mixed two kinds of the units. The lattice parameters of the other metastable stoichiometric alloys are listed in the Table **S2**. Although these compounds have comparatively higher formation enthalpies, they can provide some references for the investigations on the physical properties of Cu-Li intermetallic compounds as a function of Cu concentration. Considering that both $Cu_4Li$ and $Cu_2Li_3$ were believed to exist possibly in the previous work[6-11], we searched the structures for $Cu_4Li$ and $Cu_2Li_3$ up to 25 atoms in a unit cell. However, we found that $Cu_4Li$ and $Cu_2Li_3$ are thermodynamically unstable at atmospheric pressure. In fact, we want to point out that the minimum formation enthalpy of Cu-Li alloys is about -0.069 eV/atom, which is much higher than that of Au-Li alloys (-0.653 eV/atom).The comparatively high formation enthalpy may be the reason that the previous experimental work cannot synthetize the ordered Cu-Li alloys easily. Thus we suggest that the Cu-Li intermetallic compounds might be synthetized using some special approaches such as application of hydrostatic pressure and application of low temperature.

### 3.2 Electronic structures

To get a better understanding of the chemical bonding properties of *Fmmm* $Cu_1Li_2$, $Fd\bar{3}m$ $Cu_2Li_1$ *and* $P\bar{1}$ $Cu_7Li_1$,, the band structures, total density of states (TDOS) and partial density of states (PDOS) of the three structures were calculated, which are shown in Fig.3(a), (b) and (c) respectively. In Fig.3, the dashed red lines all correspond to the Fermi level, which was assigned at 0 eV.

As can be seen from Fig.3, for all three phases, there are band overlaps between the valence band and the conduction, indicating that they all exhibit normal metallic behaviors. Furthermore, Cu-*3d* states of the three phases are the highest in the composition of valence states and mostly responsible for bond formations. For *Fmmm* $Cu_1Li_2$, besides Cu-3d states, Li-*2p* and Cu-*4p* states also make considerable contributions near the Fermi level, while Li-2s states only show very small contribution. For $Fd\bar{3}m$ $Cu_2Li_1$ *and* $P\bar{1}$ $Cu_7Li_1$, the states of copper are more dominant than those of lithium, and Cu-*3d* and Cu-*4p* states show prominent contributions at the Fermi level.

In order to reflect directly their bonding nature, we further investigated the electron density difference of three stable Cu-Li compounds, which is defined as: $\Delta\rho = \rho_{sc} - \rho_{atom}$, where $\rho_{sc}$ is the total charge density obtained after self-consistent calculations, and $\rho_{atom}$ is the total charge density obtained after non-self-consistent calculations. Although three stable phases are metallic, the covalent bonds can be found in these structures. For *Fmmm* $Cu_1Li_2$ as shown in Fig.4(a), the co-planar Cu atoms form covalent bonds, and lithium atoms between the Cu-layers also possess some degrees of covalent character. For $Fd\bar{3}m$ $Cu_2Li_1$ in Fig.4(b), the distances between the lithium atoms are relatively large, thus they cannot form covalent bonds effectively. However, the Cu-Cu bonds still remain covalent nature, the electron transformations are mainly concentrated around tetrahedron units. $P\bar{1}$ $Cu_7Li_1$ are similar with $Cu_1Li_2$ and $Cu_2Li_1$, the covalent character is also found among Cu-Cu bonds.

To further investigate the characteristics of chemical bonding in the stable Cu-Li phases, we also performed the Mulliken atomic population and Mulliken overlap population calculations, which are shown in **Table 2** and **Table 3**, respectively. Mulliken bond population calculation is a very useful tool for evaluating the bonding character in a material. It is acknowledged that a positive value of the bond population indicates a covalent bond, and a negative value indicates an anti-bonding state, while a zero value implies a perfect ionic bond. For all three phases, the charge transfers from Li to Cu are observed as seen from **Table 2**, which indicate the ionic character in the internal of Cu-Li. Furthermore, all the Mulliken bond populations of the Cu-Li bonds have negative values as shown in **Table 3**, which also reveal the ionic character between Cu and Li atoms originated from the electronegativity difference. However, the Mulliken bond populations of Cu-Cu bonds are positive, so covalent bonds are found between copper atoms, in agreement with the analysis of the electron density difference maps.

### 3.3. Phonons and dynamical stability

Although *Fmmm* $Cu_1Li_2$, $Fd\bar{3}m$ $Cu_2Li_1$ and $P\bar{1}$ $Cu_7Li_1$, phases are thermodynamically stable at ambient condition, one phase that can exist stably must be dynamically stable. Therefore, we calculated the phonon dispersion curves, total and partial phonon density of states of *Fmmm* $Cu_1Li_2$, $Fd\bar{3}m$ $Cu_2Li_1$ and $P\bar{1}$ $Cu_7Li_1$, which are displayed in Fig.5(a), (b) and (c),respectively. In Fig.5, the absence of any imaginary phonon frequencies in the entire Brillouin zone tells us that all the three phases are dynamically stable at 0 GPa and 0K.

From the phonon curves of all three stable phases, we can see that the phonon branches all degenerated at G point. Due to the mass of Cu atom is much heavier than that of Li atom, lattice vibrations in the low frequency range derive mainly from the Cu atoms, while the contributions to high frequency vibrations mostly originates from Li atoms. For *Fmmm* $Cu_1Li_2$ shown in Fig.5(a), there are 3 acoustic modes and 24 optical modes in the full phonon dispersion, and the vibrations of Cu and Li atoms in *Fmmm* $Cu_1Li_2$ are stronger coupled. The phonon dispersion curves of $Fd\bar{3}m$ $Cu_2Li_1$ are divided into two parts, and a clear band gap between two parts is found in the phonon dispersions. The phonon density of states in Fig.5(b) indicates that the vibration modes from Cu and Li atoms are coupled to some degree. For the phonon curves of $P\bar{1}$ $Cu_7Li_1$ phase as shown in Fig.5(c), there are 3 acoustic modes and 21 optical modes. It is worth mentioning that there is a larger band gap whose value is about 4.6 THz. From the phonon DOS, we found that the vibrations of Cu and Li atoms in $P\bar{1}$ $Cu_7Li_1$ are almost independent. The high frequency vibrations located at 10-14 THz are nearly dominated by Li atoms, while the vibrations of low frequency mainly depend on Cu atoms.

### 3.4 Elastic properties

The elastic constants can provide much valuable information of a material directly or indirectly, which involves many different properties such as structure stability, brittleness, ductility, hardness, anisotropy and propagation of elastic waves. Hence, the study of elastic constants is very meaningful for a full-scale investigation of the mechanical properties of Cu-Li alloys. We calculated the elastic constants using a stress-strain method [35]. All the calculated elastic constants of Cu-Li compounds are shown in Table 4. We first investigated the mechanical stabilities of all the newly found structures. The mechanical stability of intermetallic compounds requires that strain energy must be positive, which requires that all the principal minor determinants of the elastic constant matrix $C_{ij}$ should be all positive, the detail formulas can be found in the Ref [36]. With the help of this criterion, by simple calculations, we revealed that Cu−Li compounds mentioned above are all mechanically stable at the ground state

We further calculated the bulk modulus (B), shear modulus (G), and Young's modulus (E) from the elastic constants. Theoretically, bulk modulus (B) is a measure of resistance to volume change, shear modulus (G) is a measure of resistance to reversible deformations upon shear stress[37,38], while Young's modulus (E) is often used to describe the stiffness property in intermetallic compounds. The larger the values of B and G, the stronger are the abilities of resisting to the volume and shear deformations, respectively, while the larger value of E, the stiffer is the material [37,39,40]. We calculated the bulk modulus (B) and shear modulus (G) by using the Voigte-Reusse-Hill method [41], respectively. The specific formulas for all the structures can be expressed as following:

$$B = 0.5*(B_V + B_R) \quad (2)$$

$$G = 0.5*(G_V + G_R) \quad (3)$$

where the subscript V represents the Voigt bounds [42],and R denotes the Reuss bounds[43].The values of $B_V$,$B_R$,$G_V$ and $G_R$ in Cu-Li compounds can be obtained using elastic stiffness constants $C_{ij}$ and elastic compliance coefficients $S_{ij}$[44,45]. The Young's modulus (E) was computed using the relationship [41]:

$$E = 9BG / (3B + G) \quad (4)$$

Fig.6 shows the three moduli as a function of copper content. For the Cu-Li alloys, the bulk modulus B, shear modulus G and Young's modulus E are in the range of 13.6-132.6 GPa, 5.4-48.8 GPa, 14.5-126.7 GPa, respectively. Since we have predicted so many binary Cu-Li intermetallic compounds, we can get simple relations between the elastic moduli and the copper content. The data of the three moduli (B, G and E) verse Cu content were fitted, respectively. The obtained fitting formulas are listed as following:

$$y_B=12.1+25.3x+97.9x^2 \quad (5)$$

$$y_G =7.4-26.6x+ 168.1x^2 -102.9x^3 \quad (6)$$

$$y_E = 19.3-66.6*x +422.9*x^2 -251.1*x^3 \quad (7)$$

where $x$, $y_B$, $y_G$ and $y_E$ are the copper content, bulk modulus, shear modulus and Young’s modulus of Cu-Li intermetallic compounds, respectively. As Cu content increases, the Young's modulus E increases most rapidly, then the bulk modulus B, and finally the shear modulus G. According to the previous electronic structure analysis, copper atoms trend to form covalent bonds, and the increasing Cu content results in more covalent bonds formed in the Cu-Li intermetallic compounds. Theoretically, more covalent bonds formed in a crystal are favorable for the increase of the elastic modulus, so it is natural that the three elastic moduli increase with Cu-content increasing.

Poisson's ratio is another fundament physical quantity of the ordered alloy, which is defined as the ratio of transverse strain to axial strain [46]. The Poisson’s ratio $\nu$ can be computed using the relationship [41]:

$$\nu = (3B - 2G)/(6B + 2G) \qquad (8)$$

Theoretically, Poisson's ratio is bound between -1 and 0.5 for isotropic elastic materials. The bigger Poisson’s ratio generally means the better plasticity [47]. To confirm the brittle and ductile properties of Cu-Li alloys, we also calcualted the ratio of shear modulus to bulk modulus [37]. The reference value is 0.57, which is used to separate brittle and ductile nature [39]. If G/B< 0.57, the material is ductile and vice versa. In addition, the lower the G/B ratio, the better the ductility [48-49]. Fig.7 shows that the Poisson’s ratios for Cu-Li system are in the range of 0.25 and 0.35, so all the predicted Cu-Li compounds behave in a ductile manner. Specially, the Poisson's ratios for *Fmmm* $Cu_1Li_2$, $Fd\bar{3}m$ $Cu_2Li_1$ and $P\bar{1}$ $Cu_7Li_1$ are 0.324, 0.303, and 0.324, respectively. From Fig.7, we can see that none of the G/B values of Cu-Li compounds is greater than 0.57, in agreement with the results of Poisson's ratio. Thus we can conclude that all the Cu-Li compounds are ductile materials.

We further estimated the hardness of the Cu-Li compounds using a simple relation [50]:

$$H_V = 0.1475G \qquad (9)$$

Fig.8 shows the relationship between copper content and Vickers hardness $H_v$ in Cu-Li compounds. It is found that the hardness values of the Cu-Li alloys are in the range of 0.8-7.2 GPa, which are very small on the whole. The overall trend for $H_v$ is to increase with increasing Cu content. According to equation (9), the hardness values of *Fmmm* $Cu_1Li_2$, $Fd\bar{3}m$ $Cu_2Li_1$ and $P\bar{1}$ $Cu_7Li_1$ are found to be 1.7, 5.5, and 6.6 GPa, respectively. Similar with the elastic moduli, the increase of copper content leads to more covalent bonds formed in the Cu-Li intermetallic compounds, which may be responsible for the increase of $H_v$ with copper content.

Elastic anisotropy of crystalline materials plays a very important role in various applications including phase transformations[51], anisotropic plastic deformation[52], crack behavior[53], and so on. Many different ways are widely used to characterize the anisotropy of crystal materials, for instance, the universal anisotropic index ($A^U$), the compression and shear percent anisotropies ($A_B$ and $A_G$ ). $A^U$, $A_B$ and $A_G$ can be determined via the moduli [54,55]. $A^U = 0$ represents locally isotropic materials, while $A^U$>0 denotes the extent of material anisotropy. Furthermore, the larger the value of $A^U$ is, the higher the degree of elastic anisotropy in the material is. For $A_B$ and $A_G$, a value of zero represents elastic isotropy and a value of one is the largest anisotropy.

The following equations have been used to calculate $A^U$, $A_B$ and $A_G$.

$$A^U = 5G_V / G_R + B_V / B_R - 6 \quad (10)$$

$$A_B = (B_V - B_R) / (\mathrm{B_V} + \mathrm{B_R}) \times 100\% \quad (11)$$

$$A_G = (G_V - G_R) / (G_\mathrm{V} + G_\mathrm{R}) \times 100\% \quad (12)$$

**Fig**.9 shows that $A^U$, $A_B$ and $A_G$ change with Cu content. It is found that $A^U$>0 for all Cu-Li compounds, therefore all the Cu-Li alloys exhibit anisotropic elasticity to a certain extent. The $A^U$ values of *Fmmm* $Cu_1Li_2$, $Fd\bar{3}m$ $Cu_2Li_1$ and $P\bar{1}$ $Cu_7Li_1$ are 1.91, 0.02 and 1.69, respectively, thus both *Fmmm* $Cu_1Li_2$ and $P\bar{1}$ $Cu_7Li_1$ exhibit a larger anisotropy, while $Fd\bar{3}m$ $Cu_2Li_1$ shows nearly elastic isotropy. We noticed that the percentage elastic anisotropy in compression ($A_B$) is smaller than that in shear ($A_G$) for each of the predicted Cu-Li compounds. The $A_B$ and $A_G$ values of *Fmmm* $Cu_1Li_2$ are 0.06 and 0.16, respectively, while those of $P\bar{1}$ $Cu_7Li_1$ are 0.0 and 0.15, respectively. Evidently, the shear moduli for *Fmmm*

$Cu_1Li_2$ and $P\bar{1}$ $Cu_7Li_1$ are both anisotropic, and the bulk modulus of *Fmmm* $Cu_1Li_2$ shows weaker anisotropy, while the bulk modulus of $P\bar{1}$ $Cu_7Li_1$ is isotropic. For cubic phase $Fd\bar{3}m$ $Cu_2Li_1$, both the calculated $A_B$ and $A_G$ are closed to 0, indicating that bulk modulus and shear modulus of $Cu_2Li_1$ are nearly isotropic.

**3.5 Debye temperature**

The Debye temperature $\mathbf{\Theta_D}$ is an appropriate parameter to describe some phenomena of solid-state physics, which are associated with specific heat, stability of lattices and melting points. Moreover, the $\mathbf{\Theta_D}$ is often used to estimate the strength of covalent bonds in solids. It therefore makes sense to calculate the $\mathbf{\Theta_D}$ values of Cu-Li compounds. On the other hand, as far as we know, there is temporarily no experimental data of the $\mathbf{\Theta_D}$ of Cu-Li compounds. The calculation results may provide some references for the experimental study of Cu-Li compounds in the future. We took one semi-empirical formulas to calculate the Debye temperature which is closely related to the average sound velocity. The formula adopted is given in detail as follows [56]:

$$\Theta_D = (h/k)[(3\mathrm{n}/4\pi)(\mathrm{N_A}\,\rho/M)]^{1/3} v_m \quad (13)$$

where h, k and n are the Planck's constant, Boltzmann's constant, and number of atoms per formula unit, respectively. $N_A$, $\rho$ and M are the Avogadro constant, density and molecular weight , respectively.

The average sound velocity $v_m$ can be obtained from the longitudinal and shear sound velocities ($v_l$ and $v_s$) via the following equation:

$$v_m = [(1/3)(2/v_s^3 + 1/v_l^3)]^{-1/3} \qquad (14)$$

where $v_l$ is closely related to the elastic moduli and density, and $v_s$ is determined by shear modulus G and density $\rho$.The formulas related to the $v_l$ and $v_s$ are as follows[57]:

$$v_l = [(3B + 4G)/3\rho]^{1/2} \qquad (15)$$

$$v_s = (G/\rho)^{1/2} \qquad (16)$$

Table 5 shows the values of $v_m, v_l, v_s$ and Debye temperatures at 0 K and 0 GPa. On the whole, the velocities of Cu-Li compounds increase oscillatorily with Cu-content increasing, so do Debye temperatures. The Debye temperatures of Cu-Li compounds vary in the range from 259.9K to 366.7K. Generally, the larger the Debye temperature is, the stronger the covalent bonds are [58]. As we have mentioned, Cu atoms trend to form the covalent bonds in the Cu-Li alloys, so the increasing Cu content will lead to more covalent bonds formed, and hence the strength of covalent bonds increases with Cu content on the whole. The change of Debye temperature with Cu content just proves above empirical law. For the three stable compounds, As shown in Table 5, the Debye temperature of $Fd\bar{3}m$ $Cu_2Li_1$ is slightly higher than Debye temperatures of *Fmmm* $Cu_1Li_2$ and $P\bar{1}$ $Cu_7Li_1$.This means that the covalent bonds in $Cu_2Li_1$ compound are slightly stronger than those in $Cu_1Li_2$ and $Cu_7Li_1$. Concerning Debye temperature, there is also another rule of thumb, a greater Debye temperature generally means a larger associated thermal conductivity[59]. Thus as the most stable phase, $Cu_2Li_1$ should possess the best thermal conductivity relative to the other two stable Cu–Li compounds.

## 4. Conclusion

In this paper, to resolve a longstanding controversy concerning the existence of the Cu-Li intermediate phases, we have investigated the stability of Cu-Li intermetallic compounds using the first-principles calculations coupled with the variable-composition evolutionary algorithm in USPEX. The correlations between Cu concentration and the overall performances of Cu-Li compounds have been studied systematically. The main research results are summarized as follows:

(1) We revealed that the formation enthalpy is negative for the Cu-Li intermetallic compounds at zero temperature and zero pressure, indicating the Cu-Li intermediate compounds can be synthesized at ambient condition. Three Cu-Li intermediate compounds, *Fmmm* $Cu_1Li_2$, $Fd\bar{3}m$ $Cu_2Li_1$ and $P\bar{1}$ $Cu_7Li_1$, are found to be stable in the view of thermodynamics, dynamics and mechanics.

(2) Although the three stable alloys exhibit the metallic character, covalent bonds between copper atoms and ionic bonds between lithium-copper atoms are found in these compounds. Furthermore, 3d states of copper atoms play an important role in the Cu-Li alloy and are mostly responsible for bond formations.

(3) The bulk modulus B, shear modulus G, Young's modulus E, Vicker hardness $H_v$ of the Cu-Li alloys are in the range of 13.6-132.6 GPa, 5.4-48.8 GPa, 14.5-126.7 GPa and 0.8-7.2 GPa, respectively. More importantly, B, G, E and $H_v$ of Cu-Li compounds are found to be related to Cu concentration. Cu atoms are found to trend to form the covalent bonds in the Cu-Li alloys, the increasing Cu content leads to more covalent bonds formed. So with an increase of copper content, the three elastic moduli, $H_v$ and Debye temperature increase on the whole. In addition, the fitting formulas which describe the relationships between Cu content and elastic moduli are given in detail.

(4) The Poisson's ratios for Cu-Li system are in the range of 0.25-0.35, and most Cu-Li compounds are found to have good ductility.

(5) The percentage elastic anisotropy in compression ($A_B$) is smaller than that in shear ($A_G$) for all the predicted Cu-Li compounds.

(6) The velocities and Debye temperatures increase with Cu content increasing as a whole. The Debye temperatures of Cu-Li compounds vary in the range from 259.9 K to 366.7K. As the most stable phase, $Fd\bar{3}m$ $Cu_2Li_1$ has the highest Debye temperature relative to the other stable Cu–Li compounds.

The current investigations provide important information for the Cu−Li intermetallic compounds, which will stimulate the future experiments on the structural, mechanical and thermodynamic properties measurements.

## Acknowledgement

This work is supported in China by the National Natural Science Foundation of China (Grant Nos. 51501093, 41773057,11364009, U1304612, and U1404608), Science Technology Innovation Talents in Universities of Henan Province (No.16HASTIT047), Young Core Instructor Foundation of Henan Province (No. 2015GGJS-122).

## Supporting Information

The calculated lattice parameters together with experimental values for fcc Cu and bcc Li, and the lattice parameters for the other predicted Cu-Li compounds except for $Cu_1Li_2$,$Cu_2Li_1$ and $Cu_7Li_1$.

**Table.1** The calculated lattice parameters (in Angstroms) and atomic positions for $Cu_1Li_2$, $Cu_2Li_1$ and $Cu_7Li_1$ at 0 GPa and 0K

| | Space group | Lattice parameters | Atom(site) | Wyckoff x | y | Positions z |
|---|---|---|---|---|---|---|
| $Cu_1Li_2$ | *Fmmm* | a= 13.627 | Cu(8c) | -0.50000 | -0.25000 | -0.25000 |
| | | b= 8.757 | Cu(4b) | -0.50000 | -0.50000 | -0.50000 |
| | | c= 4.856 | Li(8g) | -0.09822 | -0.50000 | -0.50000 |
| | | | Li(16o) | -0.66557 | -0.34103 | -0.50000 |
| $Cu_2Li_1$ | $Fd\bar{3}m$ | a=b=c= 6.945 | Cu(48f) | -0.37500 | 0.87500 | -0.62500 |
| | | | Li(32e) | -0.25000 | 0.75000 | -0.25000 |
| $Cu_7Li_1$ | $P\bar{1}$ | a=4.482 | Cu(2i) | 0.25514 | 1.00163 | 0.62616 |
| | | b=4.458 | Cu(2i) | 0.49621 | 0.49934 | 0.24916 |
| | | c=5.114 | Cu(2i) | 0.24841 | 1.00687 | 0.12230 |
| | | $\alpha$ =73.3 | Cu(1c) | 0.00000 | 0.50000 | -0.00000 |
| | | $\beta$ =90.0 | Li(1g) | 0.00000 | 0.50000 | 0.50000 |
| | | $\gamma$ =99.1 | | | | |

**Table.2** Mulliken atomic population analysis of $Cu_1Li_2$, $Cu_2Li_1$ and $Cu_7Li_1$

| Compound | Species | s | p | d | Total | Charge(e) |
|---|---|---|---|---|---|---|
| | $Cu_{8c}$ | 0.88 | 1.35 | 9.80 | 12.03 | -1.03 |
| $Cu_1Li_2$ | $Cu_{4b}$ | 0.98 | 1.36 | 9.82 | 12.16 | -1.16 |
| | $Li_{8g}$ | 2.26 | 0.00 | 0.00 | 2.26 | 0.74 |
| | $Li_{16o}$ | 2.56 | 0.00 | 0.00 | 2.56 | 0.44 |
| | $Cu_{48f}$ | 0.65 | 1.36 | 9.77 | 11.78 | -0.78 |
| $Cu_2Li_1$ | $Li_{16c}$ | 1.45 | 0.00 | 0.00 | 1.45 | 1.55 |
| | $Cu_{2i}$ | 0.55 | 1.03 | 9.74 | 11.32 | -0.32 |
| | $Cu_{2i}$ | 0.51 | 0.91 | 9.73 | 11.15 | -0.15 |
| $Cu_7Li_1$ | $Cu_{2i}$ | 0.55 | 1.00 | 9.74 | 11.29 | -0.29 |
| | $Cu_{1d}$ | 0.56 | 1.00 | 9.74 | 11.31 | -0.31 |
| | $Li_{1e}$ | 1.18 | 0.00 | 0.00 | 1.18 | 1.82 |

**Table. 3** Mulliken bond population analysis of $Cu_1Li_2$, $Cu_2Li_1$ and $Cu_7Li_1$

| Compound | Bond | Population | Length(Å) | Compound | Bond | Population | Length(Å) |
|---|---|---|---|---|---|---|---|
| | $Cu_{8c}$-$Cu_{8c}$ | 2.23 | 2.467 | | $Cu_{2i}$-$Cu_{1d}$ | 0.36 | 2.535 |
| | $Cu_{8c}$-$Cu_{4b}$ | 1.15 | 2.520 | | $Cu_{2i}$-$Cu_{2i}$ | 0.43 | 2.542 |
| $Cu_1Li_2$ | $Li_{16o}$-$Cu_{4b}$ | -0.39 | 2.612 | | $Cu_{2i}$-$Cu_{2i}$ | 0.41 | 2.549 |
| | $Li_{16o}$-$Cu_{8c}$ | -0.32 | 2.661 | | $Cu_{2i}$-$Cu_{2i}$ | 0.20 | 2.552 |
| | $Li_{8g}$-$Cu_{4b}$ | -0.44 | 2.805 | $Cu_7Li_1$ | $Cu_{2i}$-$Cu_{2i}$ | 0.35 | 2.554 |
| | $Li_{8g}$-$Cu_{8c}$ | -0.24 | 2.851 | | $Cu_{2i}$-$Cu_{2i}$ | 0.39 | 2.556 |
| | $Li_{8g}$-$Li_{8g}$ | -0.01 | 2.668 | | $Cu_{2i}$-$Cu_{1d}$ | 0.48 | 2.559 |
| | $Li_{16o}$-$Li_{16o}$ | 0.15 | 2.757 | | $Cu_{2i}$-$Cu_{2i}$ | 0.00 | 2.559 |
| | $Li_{16o}$-$Li_{16o}$ | 0.28 | 2.772 | | $Cu_{2i}$-$Cu_{2i}$ | 0.16 | 2.564 |
| | $Li_{16o}$-$Li_{16o}$ | -0.17 | 2.953 | | $Cu_{2i}$-$Cu_{2i}$ | 0.38 | 2.567 |
| | $Li_{8g}$-$Li_{16o}$ | 0.30 | 2.964 | | $Cu_{2i}$-$Cu_{2i}$ | 0.53 | 2.570 |
| | | | | | $Li_{1e}$-$Cu_{2i}$ | -0.55 | 2.560 |
| $Cu_2Li_1$ | $Cu_{48f}$-$Cu_{8b}$ | 0.60 | 2.452 | | $Li_{1e}$-$Cu_{1d}$ | -0.55 | 2.567 |
| | $Li_{32e}$-$Cu_{48f}$ | -0.25 | 2.876 | | $Li_{1e}$-$Cu_{2i}$ | -0.67 | 2.571 |

**Table.4** The calculated independent elastic constants (in GPa) of the predicted intermetallics in the binary Cu-Li systems at 0K and 0GPa.

| | $C_{11}$ | $C_{22}$ | $C_{33}$ | $C_{44}$ | $C_{55}$ | $C_{66}$ | $C_{12}$ | $C_{13}$ | $C_{14}$ | $C_{15}$ | $C_{16}$ |
|---|---|---|---|---|---|---|---|---|---|---|---|
| $Cu_1Li_8$ | 22.6 | 24.0 | 18.7 | 12.6 | 13.1 | 8.1 | 9.7 | 14.4 | -1.5 | 0 | -6.1 |
| $Cu_1Li_7$ | 24.9 | 27.2 | 17.0 | 13.9 | 10.8 | 6.1 | 8.2 | 14.1 | -0.7 | -0.8 | 6.7 |
| $Cu_1Li_6$ | 27.1 | 30.6 | 25.7 | 9.1 | 12.4 | 8.2 | 12.3 | 16.3 | -0.2 | 1.9 | -3.9 |
| $Cu_1Li_5$ | 34.1 | 49.6 | 36.6 | 4.6 | 2.4 | 11.5 | 13.2 | 6.9 | | | |
| $Cu_1Li_4$ | 52.1 | 52.1 | 35.2 | 6.5 | 5.7 | 17.0 | 18.5 | 7.4 | -2.4 | 6.4 | 0.8 |
| $Cu_1Li_3$ | 39.5 | 42.8 | 24.1 | 15.1 | 20.0 | 6.4 | 13.4 | 23.7 | -0.2 | -2.0 | -0.3 |
| $Cu_1Li_2$ | 40.1 | 67.8 | 69.5 | 12.2 | 5.9 | 6.4 | 14.5 | 9.3 | | | |
| $Cu_2Li_3$ | 54.6 | | | 19.8 | | | 31.4 | | | | |
| $Cu_1Li_1$ | 111.2 | | 67.4 | 17.8 | | | 42.0 | 19.3 | | | |
| $Cu_6Li_5$ | 68.1 | 79.8 | 96.3 | 34.4 | 31.6 | 42.6 | 46.9 | 36.8 | -1.6 | -13.6 | 0.6 |
| $Cu_4Li_3$ | 109.2 | | 116.1 | 23.4 | | | 38.7 | 26.4 | | | |
| $Cu_3Li_2$ | 106.5 | 123.3 | 100.7 | 13.5 | 20.1 | 33.0 | 42.6 | 38.0 | -5.9 | -9.6 | 6.1 |
| $Cu_7Li_4$ | 98.1 | 119.9 | 111.9 | 35.4 | 54.4 | 42.5 | 49.8 | 60.0 | 5.1 | 1.9 | -17.7 |
| $Cu_2Li_1$ | 125.1 | | | 35.3 | | | 45.0 | | | | |
| $Cu_8Li_3$ | 121.1 | 110.8 | 129.0 | 56.6 | 47.2 | 60.8 | 68.2 | 53.3 | 3.3 | -15.4 | -2.7 |
| $Cu_3Li_1$ | 169.4 | 146.0 | 154.8 | 47.6 | 28.6 | 35.8 | 55.6 | 45.4 | 3.0 | 5.0 | 2.0 |
| $Cu_7Li_2$ | 155.2 | 165.6 | 160.7 | 32.6 | 42.5 | 38.1 | 54.1 | 60.9 | -16.1 | -13.3 | -12.6 |
| $Cu_4Li_1$ | 155.8 | 139.0 | 136.8 | 64.7 | 48.9 | 52.7 | 66.7 | 68.4 | 5.7 | 15.8 | -17.1 |
| $Cu_5Li_1$ | 184.8 | 184.0 | 137.2 | 67.1 | 63.6 | 18.6 | 41.7 | 88.5 | | 3.9 | |
| $Cu_6Li_1$ | 183.6 | | 198.5 | 35.6 | | | 73.7 | 55.1 | | | |
| $Cu_7Li_1$ | 171.6 | 175.6 | 189.4 | 44.0 | 49.4 | 60.4 | 86.4 | 77.3 | 5.8 | 25.3 | -7.4 |
| $Cu_8Li_1$ | 193.1 | 184.8 | 154.6 | 66.8 | 57.0 | 35.8 | 61.3 | 87.9 | 9.2 | -12.4 | 11.4 |
| $Cu_{11}Li_1$ | 209.7 | 194.5 | 195.1 | 54.1 | 39.0 | 39.0 | 67.7 | 64.2 | | 2.4 | |

| | $C_{23}$ | $C_{24}$ | $C_{25}$ | $C_{26}$ | $C_{34}$ | $C_{35}$ | $C_{36}$ | $C_{45}$ | $C_{46}$ | $C_{56}$ |
|---|---|---|---|---|---|---|---|---|---|---|
| $Cu_1Li_8$ | 14.4 | 1.8 | 0.4 | 5.2 | -0.5 | -0.6 | 0.3 | 1.1 | 0.1 | 0 |
| $Cu_1Li_7$ | 16.3 | -0.9 | 0.1 | -5.6 | 0.8 | 0.3 | -0.2 | 0.5 | 0.2 | -0.5 |
| $Cu_1Li_6$ | 13.5 | -4.2 | 1.1 | 3.6 | 4.9 | -1.0 | -0.2 | 0.4 | 1.1 | 0.1 |
| $Cu_1Li_5$ | 7.4 | | | | | | | | | |
| $Cu_1Li_4$ | 8.4 | -7.4 | -0.3 | -0.1 | 4.5 | -2.0 | -0.2 | -0.4 | -0.8 | -2.5 |
| $Cu_1Li_3$ | 14.2 | -0.2 | 1.3 | 0.5 | -0.2 | 0.4 | 1.0 | 0.2 | 0.9 | -0.2 |
| $Cu_1Li_2$ | 26.4 | | | | | | | | | |
| $Cu_2Li_3$ | | | | | | | | | | |
| $Cu_1Li_1$ | | | | | | | | | | |
| $Cu_6Li_5$ | 37.1 | -1.1 | 5.9 | -15.5 | 1.0 | 16.2 | -3.9 | -0.8 | 1.9 | -2.9 |
| $Cu_4Li_3$ | | | | | | | | | | |
| $Cu_3Li_2$ | 32.6 | -7.0 | -3.6 | -6.1 | 2.5 | 9.9 | 2.4 | 1.2 | -1.3 | -4.8 |
| $Cu_7Li_4$ | 48.4 | 7.3 | -4.4 | 18.0 | -12.1 | 7.0 | 8.3 | 4.3 | -4.5 | 3.1 |
| $Cu_2Li_1$ | | | | | | | | | | |
| $Cu_8Li_3$ | 72.9 | 1.4 | 0.7 | 13.7 | 3.7 | 21.2 | 0.7 | -0.8 | -3.0 | 1.1 |
| $Cu_3Li_1$ | 61.5 | 0.4 | 23.0 | 0.9 | 0.5 | -21.6 | 0 | 0.6 | 20.4 | 0.7 |
| $Cu_7Li_2$ | 47.5 | -6.1 | 12.2 | 12.3 | 21.2 | 5.8 | -4.8 | -3.7 | 11.4 | -14.3 |
| $Cu_4Li_1$ | 83.3 | -0.7 | 4.7 | 21.3 | -8.6 | -22.7 | -3.3 | -5.6 | 1.8 | 5.1 |
| $Cu_5Li_1$ | 90.03 | | 0.14 | | | -0.02 | | | -1.21 | |
| $Cu_6Li_1$ | | | | | | | | | | |
| $Cu_7Li_1$ | 71.1 | 10.9 | -17.1 | 15.5 | -18.7 | -10.1 | -7.8 | -9.1 | -12.5 | 1.1 |
| $Cu_8Li_1$ | 94.2 | 5.1 | -9.5 | -22.3 | -12.5 | 21.8 | 10.2 | 7.0 | -6.2 | 7.6 |
| $Cu_{11}Li_1$ | 84.7 | | 27.3 | | | -22.3 | | | 19.7 | |

**Table. 5** Density ρ (g/cm$^3$), shear sound velocity $v_s$ (m/s), longitudinal sound velocity $v_l$ (m/s),average sound velocity $v_m$ (m/s) and Debye temperature $\Theta_D$ (K) of $Cu_xLi_y$ binary alloys at 0K and 0 GPa.

| | Crystal system | $\rho$ | $v_s$ | $v_l$ | $v_m$ | $\Theta_D$ |
|---|---|---|---|---|---|---|
| $Cu_1Li_8$ | triclinic | 1.195 | 2259 | 4472 | 2533 | 285.7 |
| $Cu_1Li_7$ | triclinic | 1.290 | 2037 | 4237 | 2291 | 259.9 |
| $Cu_1Li_6$ | triclinic | 1.440 | 2223 | 4416 | 2493 | 286.8 |
| $Cu_1Li_5$ | orthorhombic | 1.475 | 2307 | 4479 | 2583 | 291.1 |
| $Cu_1Li_4$ | triclinic | 1.689 | 2439 | 4597 | 2725 | 309.8 |
| $Cu_1Li_3$ | triclinic | 2.222 | 2092 | 3985 | 2339 | 277.7 |
| $Cu_1Li_2$ | orthorhombic | 2.663 | 2089 | 4097 | 2341 | 275.9 |
| $Cu_2Li_3$ | cubic | 3.383 | 2173 | 4227 | 2434 | 296.9 |
| $Cu_1Li_1$ | hexagonal | 4.051 | 2554 | 4526 | 2840 | 347.1 |
| $Cu_6Li_5$ | triclinic | 4.848 | 2222 | 4194 | 2484 | 314.8 |
| $Cu_4Li_3$ | hexagonal | 5.089 | 2489 | 4422 | 2770 | 352.2 |
| $Cu_6Li_4$ | triclinic | 4.941 | 2226 | 4360 | 2494 | 309.9 |
| $Cu_7Li_4$ | triclinic | 5.654 | 2393 | 4506 | 2674 | 341.8 |
| $Cu_2Li_1$ | cubic | 5.325 | 2640 | 4771 | 2941 | 363.8 |
| $Cu_8Li_3$ | triclinic | 6.467 | 2441 | 4559 | 2726 | 351.0 |
| $Cu_2Li_1$ | triclinic | 6.709 | 2303 | 4497 | 2580 | 333.3 |
| $Cu_7Li_2$ | triclinic | 6.929 | 2301 | 4471 | 2577 | 333.1 |
| $Cu_4Li_1$ | triclinic | 7.125 | 2389 | 4598 | 2674 | 345.9 |
| $Cu_5Li_1$ | monoclinic | 7.427 | 2385 | 4663 | 2672 | 346.4 |
| $Cu_6Li_1$ | hexagonal | 7.606 | 2534 | 4711 | 2829 | 366.7 |
| $Cu_7Li_1$ | triclinic | 7.774 | 2391 | 4691 | 2680 | 347.8 |
| $Cu_8Li_1$ | triclinic | 7.876 | 2350 | 4663 | 2635 | 341.9 |
| $Cu_{11}Li_1$ | monoclinic | 8.104 | 2364 | 4647 | 2649 | 343.9 |

\

**Figure captions.**

**Fig 1.** Convex hull of the Cu−Li system at atmospheric pressure. The black line indicates the ground-state convex hull. The fcc-Cu and bcc-Li are used as the reference states.

**Fig.2** The unit cells of (a) $Cu_1Li_2$, (b) $Cu_2Li_1$ and (c) $Cu_7Li_1$phases which are stable for the considered Cu-Li compounds. Cu(Li) are colored red (green).

**Fig.3** Electronic band structures and densities of states of (a) $Cu_1Li_2$, (b) $Cu_2Li_1$ and (c) $Cu_7Li_1$ at ground state.

**Fig.4** The three dimensional charge density difference map for the Cu-Li alloys. (a) $Cu_1Li_2$, (b) $Cu_2Li_1$ and (c) $Cu_7Li_1$ at ground state. Cu(Li) are colored red (green).

**Fig.5** Calculated Phonon dispersions along high-symmetry directions in the Brillouin zone and densities of states of (a) $Cu_1Li_2$, (b) $Cu_2Li_1$ and (c) $Cu_7Li_{1.}$

**Fig.6** The calculated bulk modulus (B), shear modulus (G) and Young's modulus (E) as a function of Cu concentration.

**Fig.7** Calculated G/B and Poisson's rate $\nu$ of Cu−Li compounds as a function of Cu concentration.

**Fig.8** Calculated hardness $H_V$ of Cu−Li compounds as a function of Cu concentration.

**Fig.9** Calculated universal anisotropic index ($A^U$), compression and shear

percent anisotropies ($A_B$ and $A_G$) of Cu−Li compounds as a function of Cu concentration.

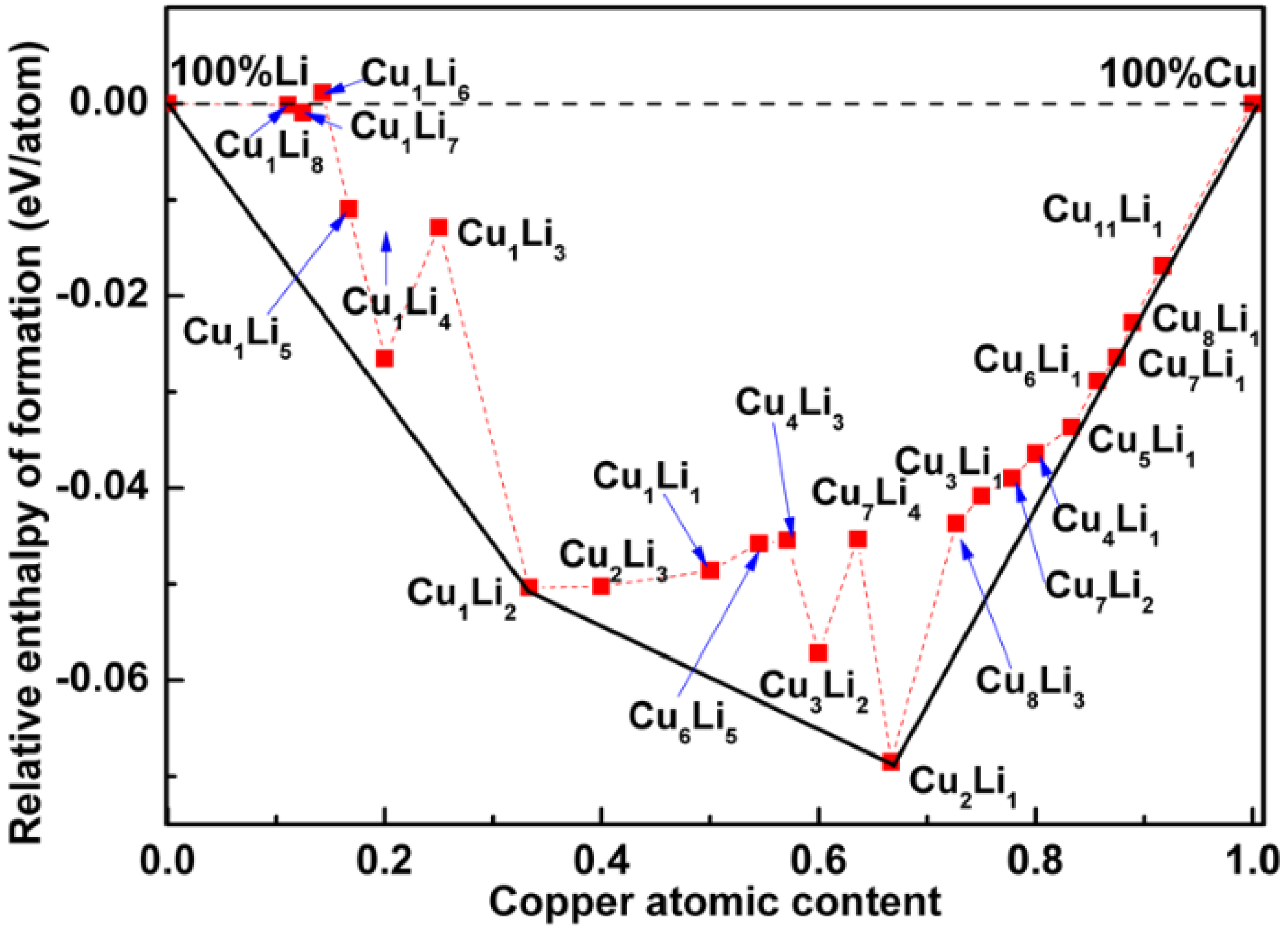

**Fig 1.** Convex hull of the Cu−Li system at atmospheric pressure. The black line indicates the ground-state convex hull. The fcc-Cu and bcc-Li are used as the reference states.

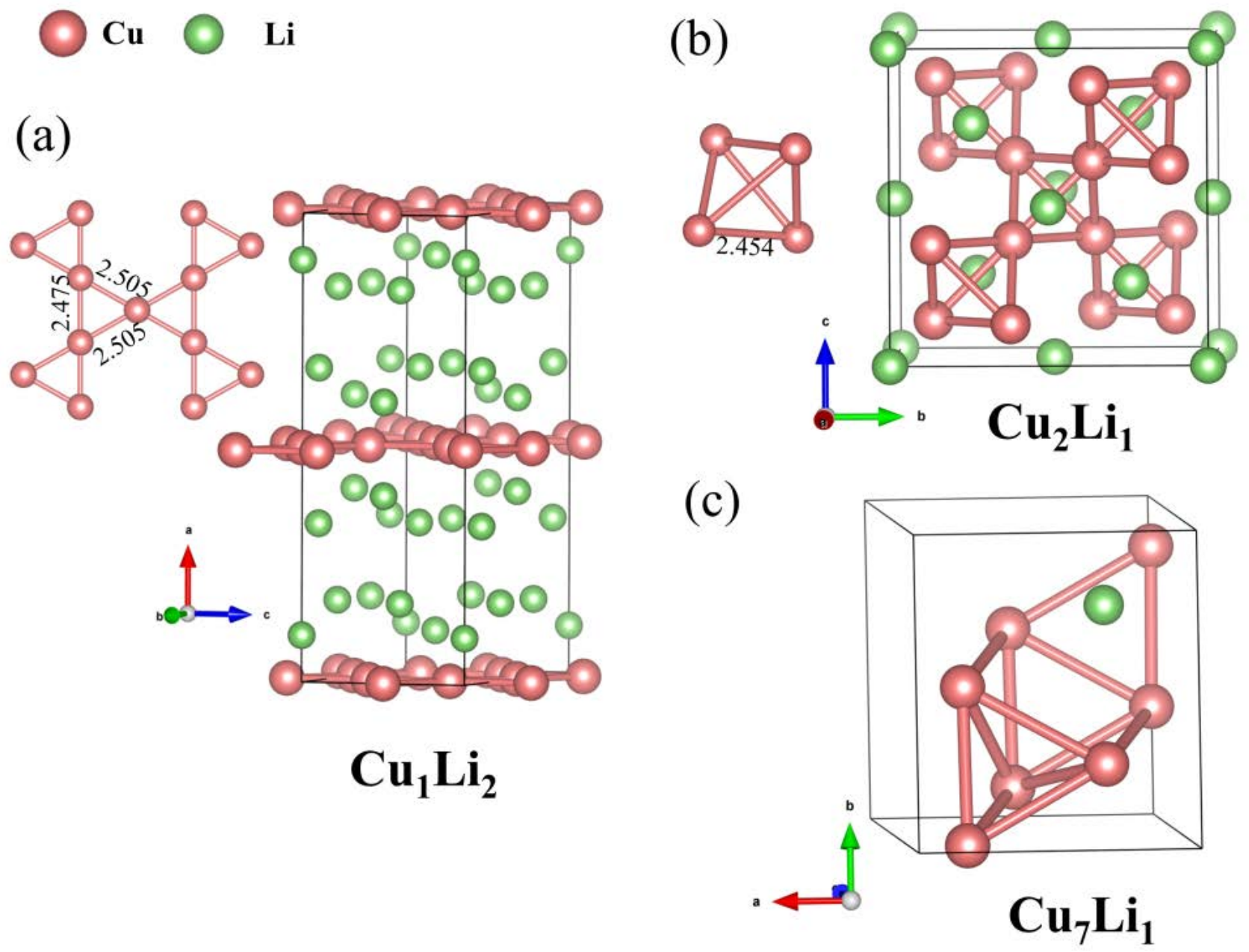

**Fig.2** The unit cells of (a) $Cu_1Li_2$, (b) $Cu_2Li_1$ and (c) $Cu_7Li_1$phases which are stable for the considered Cu-Li compounds. Cu(Li) are colored red (green).

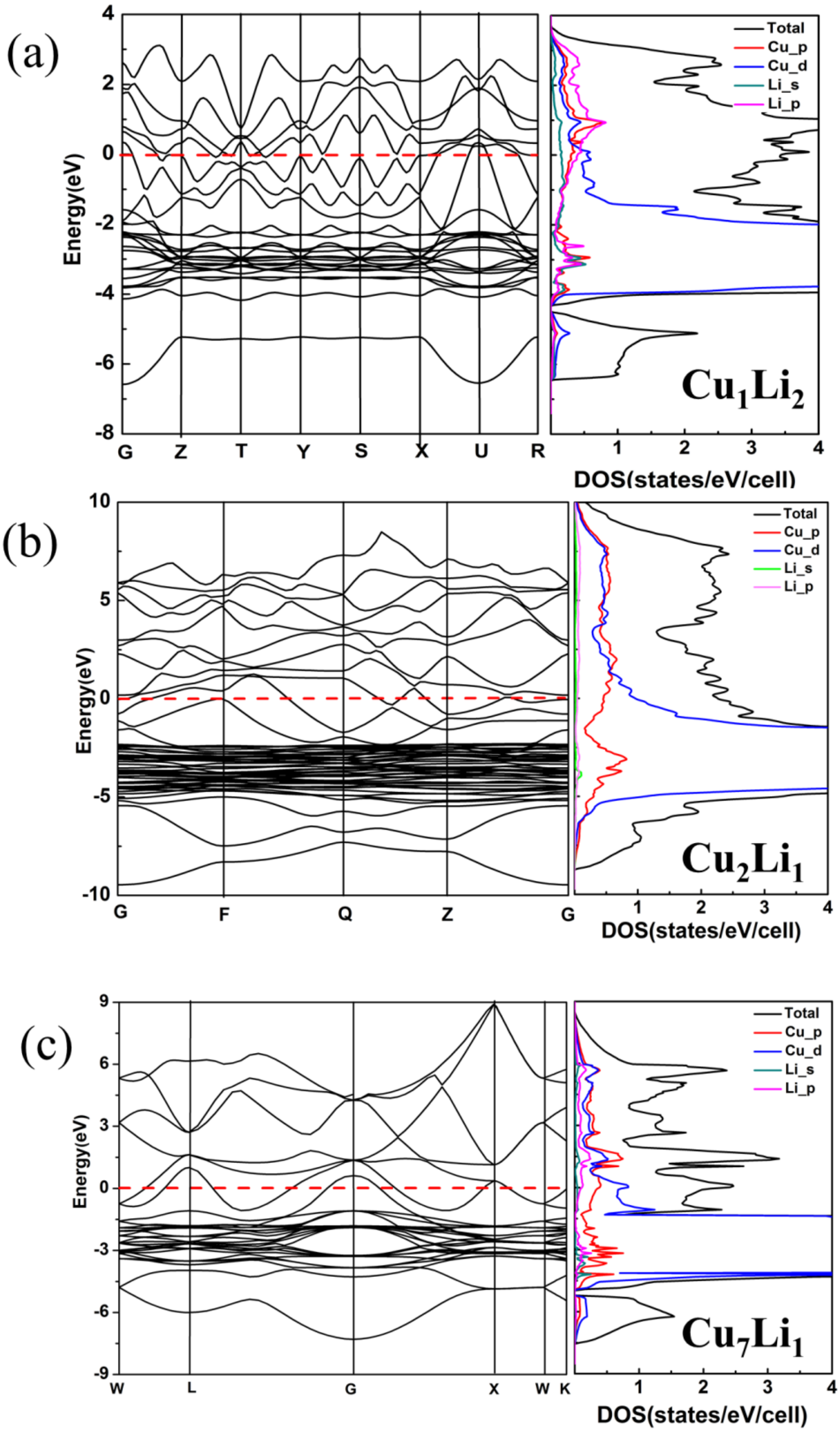

**Fig.3** Electronic band structures and densities of states of (a) $Cu_1Li_2$, (b) $Cu_2Li_1$ and (c) $Cu_7Li_1$ at ground state.

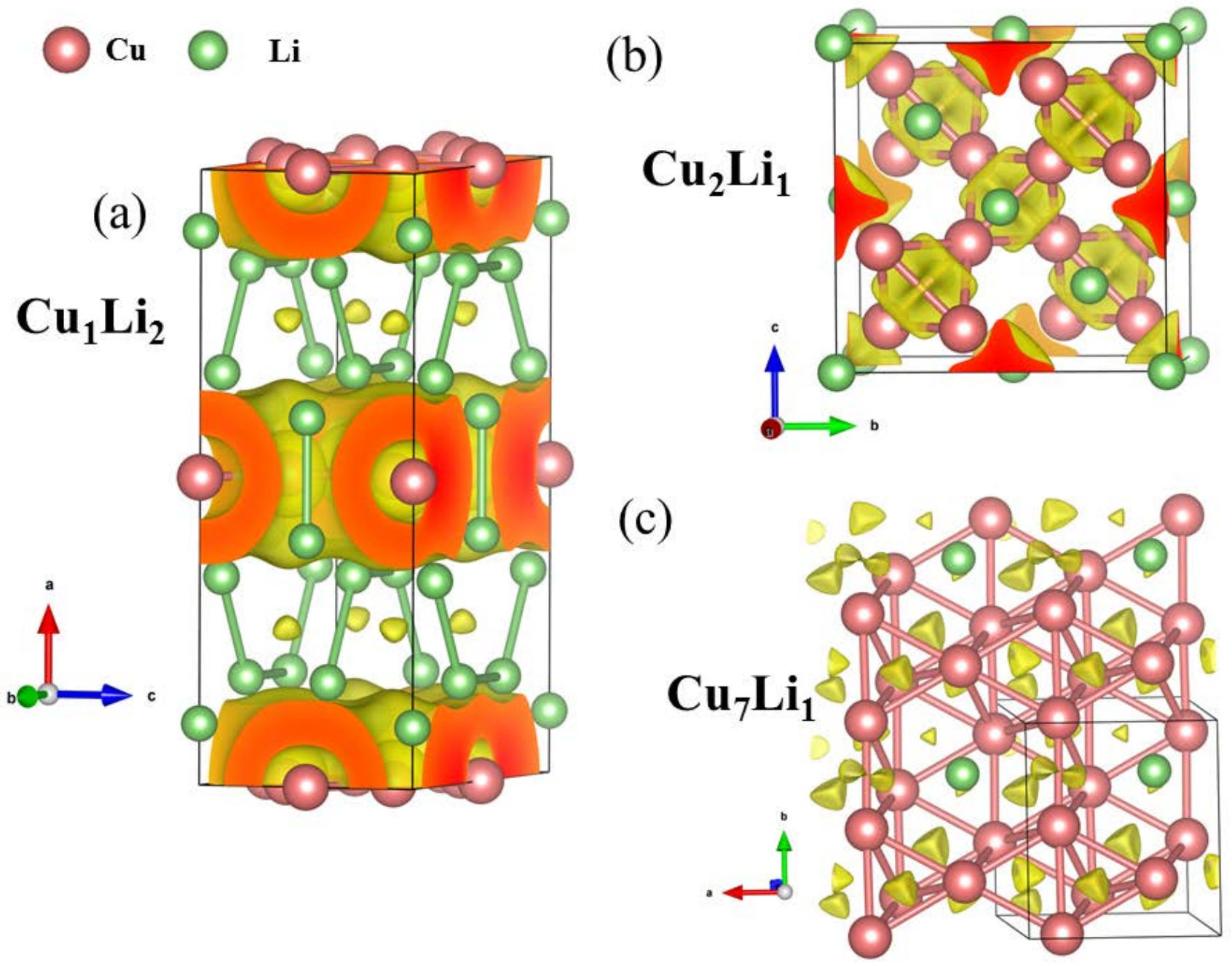

**Fig.4** The three dimensional charge density difference map for the Cu-Li alloys. (a) $Cu_1Li_2$, (b) $Cu_2Li_1$ and (c) $Cu_7Li_1$ at ground state. Cu(Li) are colored red (green).

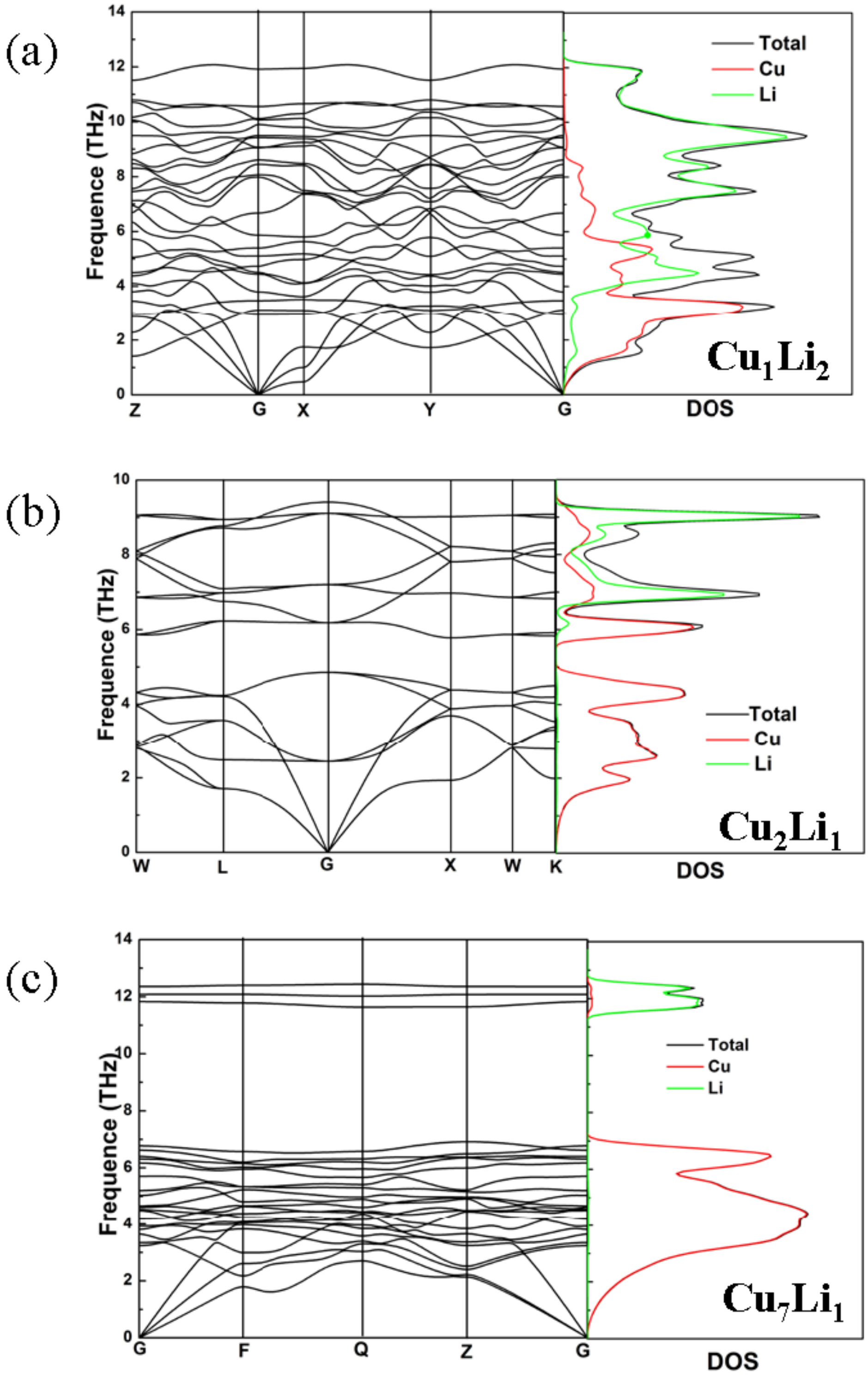

**Fig.5** Calculated Phonon dispersions along high-symmetry directions in the Brillouin zone and densities of states of (a) $Cu_1Li_2$, (b) $Cu_2Li_1$ and (c) $Cu_7Li_1$.

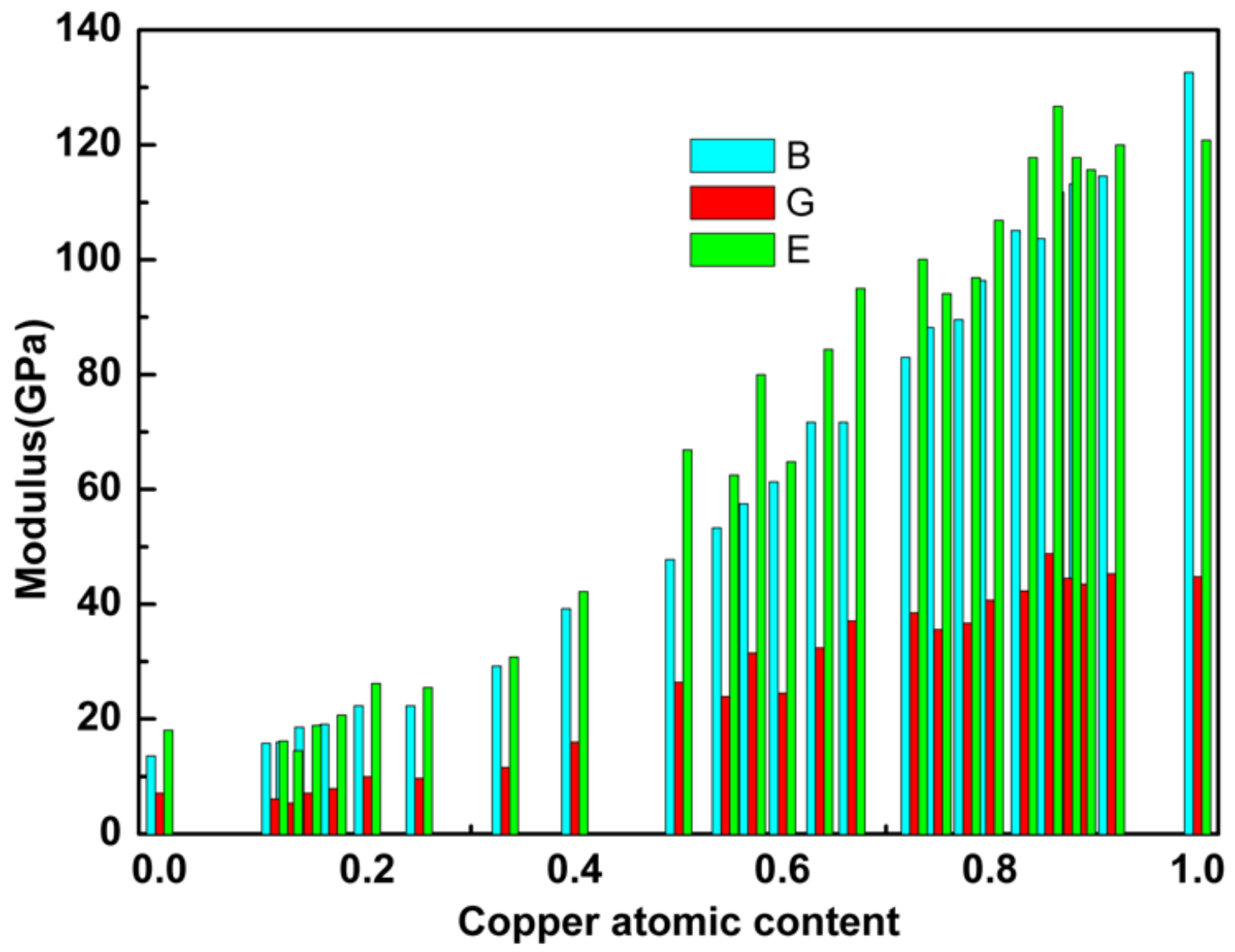

**Fig.6** The calculated bulk modulus (B), shear modulus (G) and Young's modulus (E) as a function of Cu concentration.

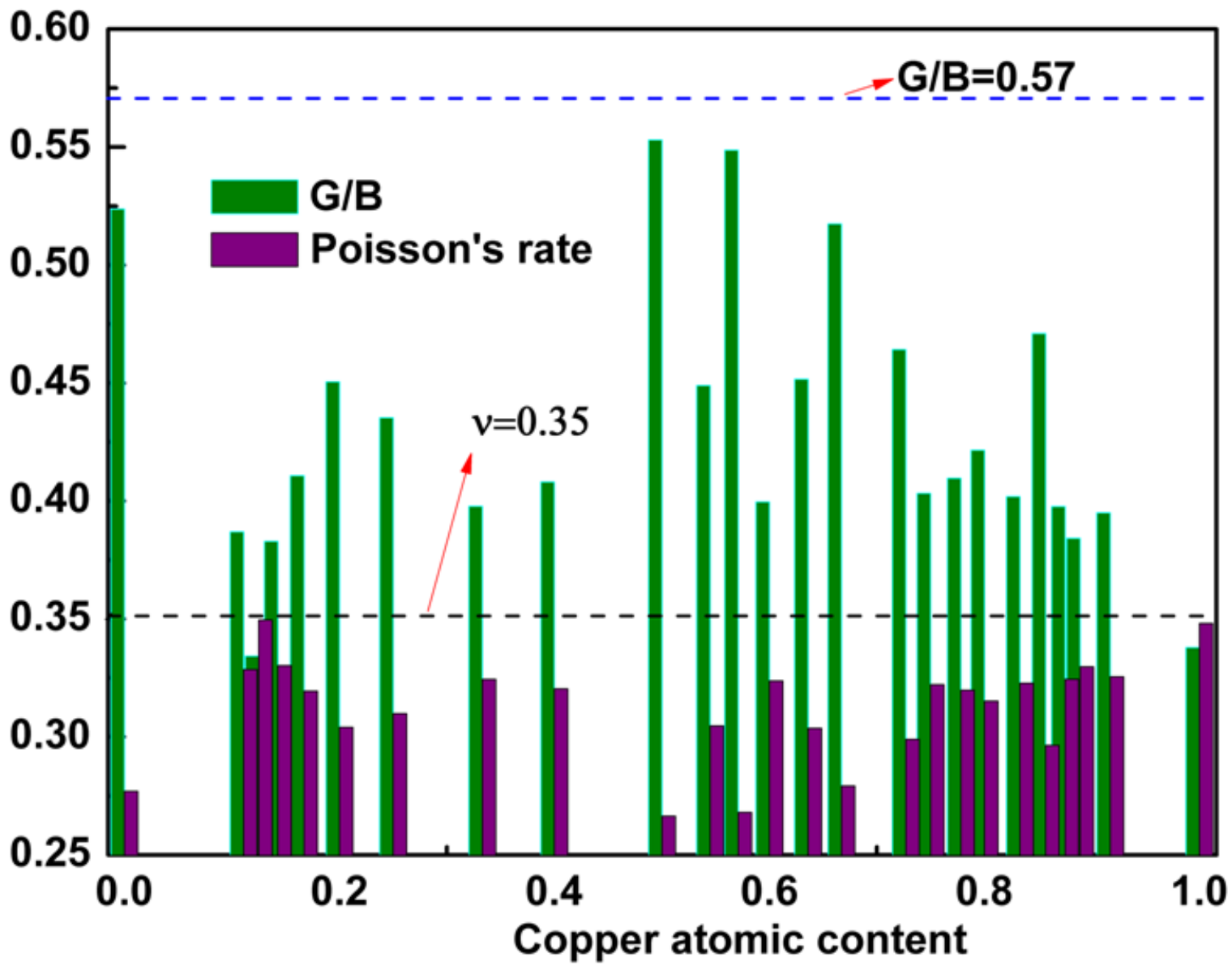

**Fig.7** Calculated G/B and Poisson's rate $\nu$ of Cu−Li compounds as a function of Cu concentration.

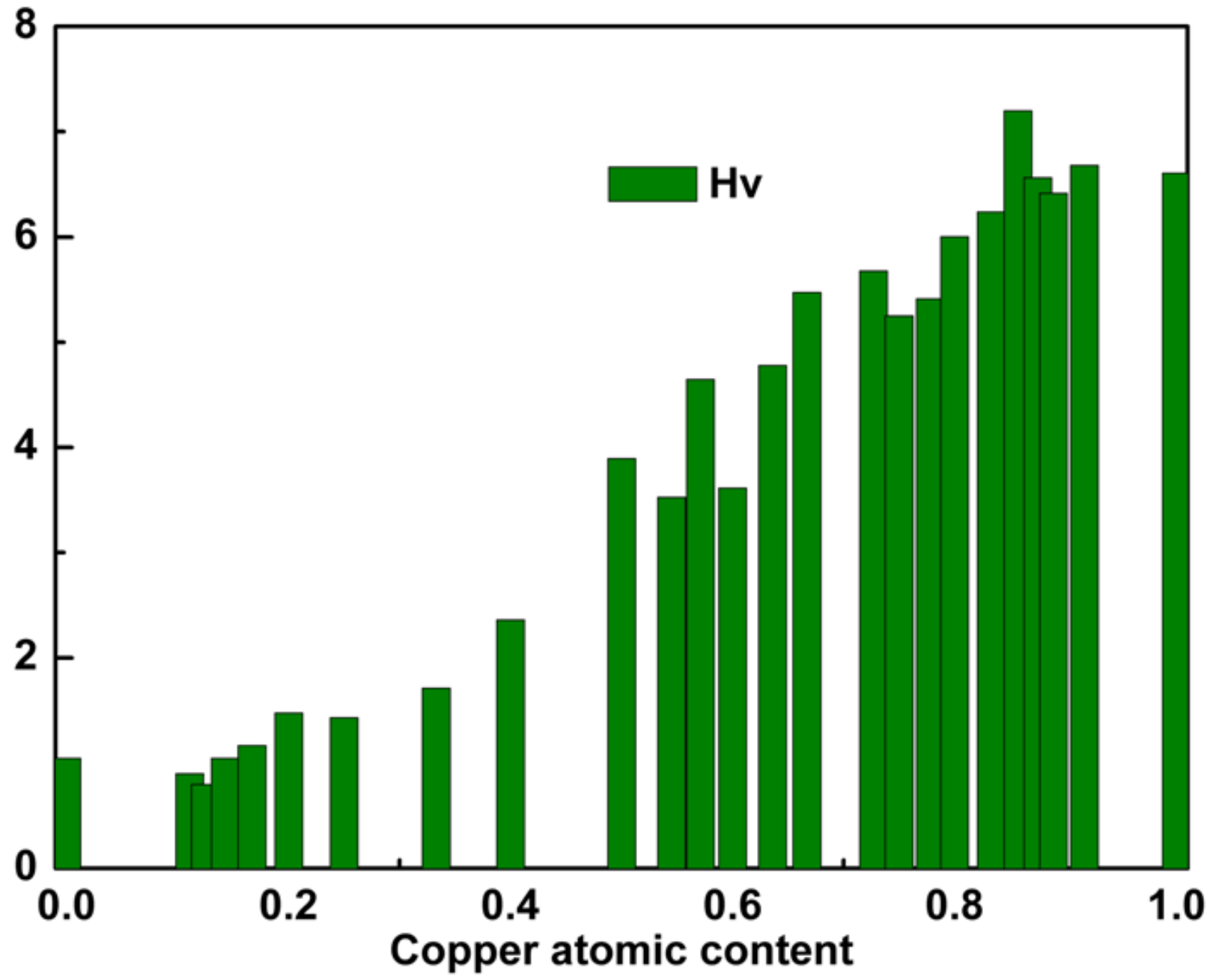

**Fig.8** Calculated hardness $H_V$ of Cu−Li compounds as a function of Cu concentration.

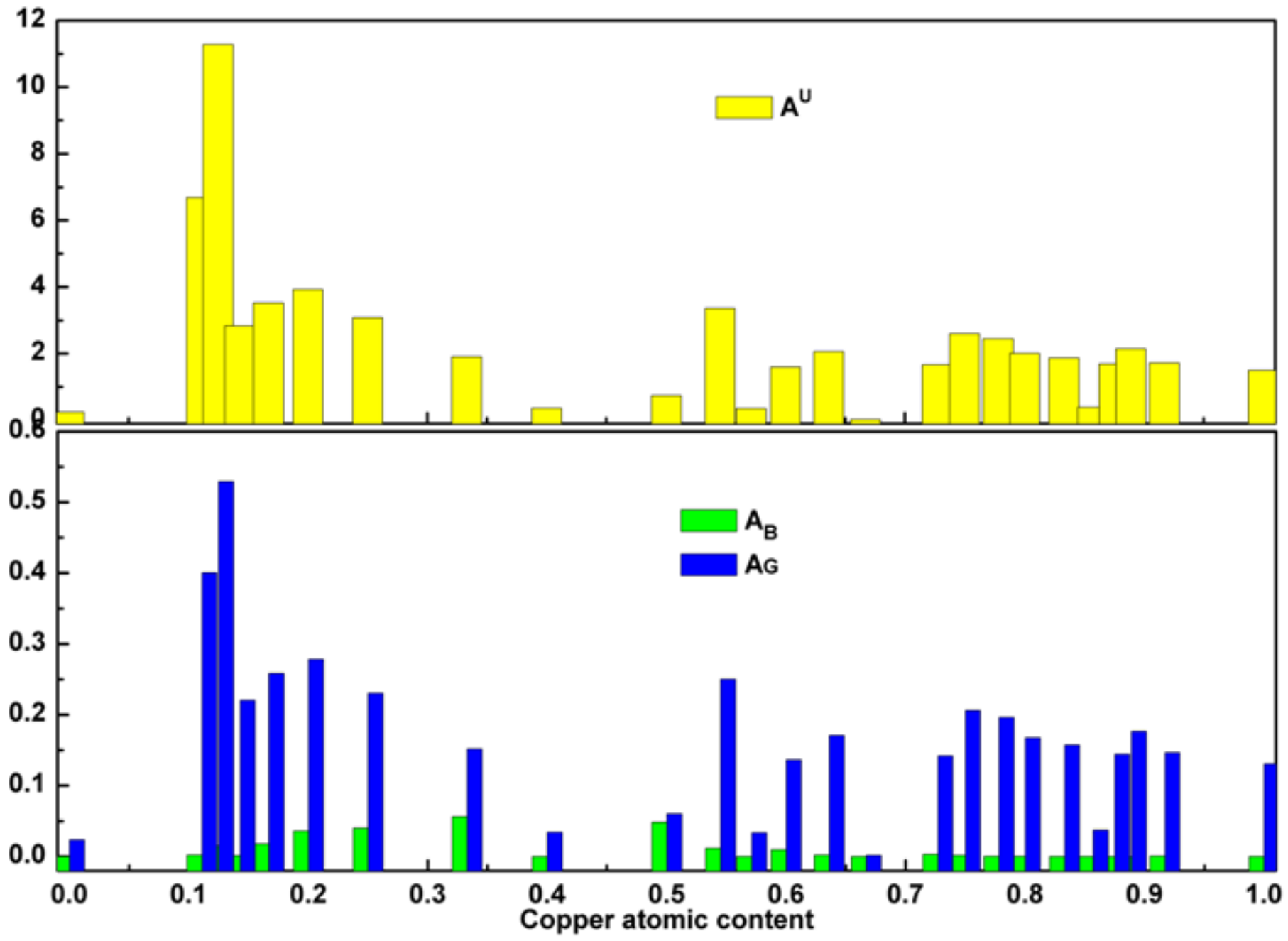

**Fig.9** Calculated universal anisotropic index ($A^U$), compression and shear percent anisotropies ($A_B$ and $A_G$) of Cu−Li compounds as a function of Cu concentration.